\documentclass[namedreferences]{kluwer}
\usepackage[dvips]{epsfig}

\begin{document}
\begin{article}
\begin{opening}

\title{INTERMEDIATE$-$TERM PERIODICITIES IN SOFT X-RAY FLARE INDEX DURING SOLAR
CYCLES 21, 22 AND 23}

\author{BHUWAN \surname{JOSHI} and ANITA JOSHI}

\institute{Aryabhatta Research Institute of Observational Sciences, 
Manora Peak, Naini Tal 263 129, Uttaranchal, India
(E-mail: bhuwan@aries.ernet.in)}

\begin{ao}
Bhuwan Joshi\\
Aryabhatta Research Institute of Observational Sciences (ARIES),\\
Manora Peak, Naini Tal 263 129,\\
INDIA.\\
Phone: +91-05942-235583, 235136\\
Fax: +91-05942-235136\\
email: bhuwan@upso.ernet.in
\end{ao}

\begin{abstract}
We have analyzed the intermediate term periodicities in soft X-ray flare index 
($FI_{SXR}$) during solar cycles
21, 22 and 23. Power spectral analysis
of daily $FI_{SXR}$ reveals a significant period of 161 days in cycle 21 which
is absent during cycle 22 and 23. We have found that in cycle 22 
periodicities of 74 and 83 days are in operation. A 123 day periodicity
has been found to be statistically significant during the part of the
current solar cycle 23. The existence of these periodicities has been
discussed in the light of earlier results.
 
\end{abstract}

\end{opening}

\section{Introduction}
In the solar cycle 21, an intermediate term (or midrange) periodicity of 154 days was 
discovered by Rieger et al. (1984) in gamma ray and soft X-ray flare data. 
Since then a number of studies have been made to look for the evidence
of near 155 days periodicity in solar activity. With solar flares and
flare related data it has been detected by Bogart and Bai (1985), 
Ichimoto et al. (1985), \"{O}zg\"{u}c and Ata\c{c} (1989), 
Dr\"{o}ge et al. (1990), Bai and Cliver (1990), and Kile and Cliver (1991).
The presence of this periodicity was also analyzed in several other
solar activity features, such as, sunspot area (Lean, 1990; Carbonell and 
Ballester, 1990, 1992), sunspot blocking function (Lean and Brueckner, 1989),
10.7 cm radio flux (Lean and Brueckner, 1989), and Photospheric magnetic 
flux (Ballester, Oliver, and Carbonell, 2002).
On the other hand no evidence for the presence of a near 155 days periodicity 
has been found in any solar activity indicator during solar cycle 22 and
part of cycle 23 (Kile and Cliver, 1991; \"{O}zg\"{u}c and 
Ata\c{c}, 1994; Oliver and Ballester, 1995; Ballester, Oliver, and 
Carbonell, 2002; Bai, 2003) instead some other prominent periodicities
were detected in these studies.

In the present paper, we have made an attempt to investigate the existence
of intermediate term periodicities in the soft X-ray flare index ($FI_{SXR}$), which is 
based on the continuous record of soft X-ray (SXR) flares observed by 
GOES during solar cycles 21, 22, and 23. The $FI_{SXR}$ is  
calculated by 
weighing  peak intensity of SXR flares of different classes (B to X) 
in units of $10^{-6}$ $Wm^{-2}$ (see Joshi and Joshi, 2004 and 
references therein). In this manner the daily values of $FI_{SXR}$ represent 
the daily SXR flare activity and are suitable for a short term solar activity
analysis. To compute the $FI_{SXR}$, the SXR flare data for the time span of 
01 June 1976 to 31 May 2004 has been downloaded from NGDC's 
anonymous ftp server: ftp://ftp.ng dc.no aa.gov/STP/
SOLAR$\_$DATA/SOLAR$\_$FLARES/XRAY$\_$FLARES. In Figure 1 the time 
variation of 27-day moving average of daily 
$FI_{SXR}$ for the period 1976$-$2004 is presented. 

To perform the study, we have split the data in three parts corresponding
to the period of solar cycle 21 (June 1976 to August 1986) with 3717 days data, 
 cycle 22
(September 1986 to April 1996) with 3524 days data and cycle 23 (May 1996 
to May 2004) with 2953 days data. 
Cycle 23 is still not complete and the data for this cycle covers the
ascending phase, the maximum and part of the descending phase of the 
solar cycle. Also, for cycle 23 we have performed the power spectra for
three different time intervals corresponding to the periods of time studied
by Bai (2003).

\begin{figure}
\epsfig{file=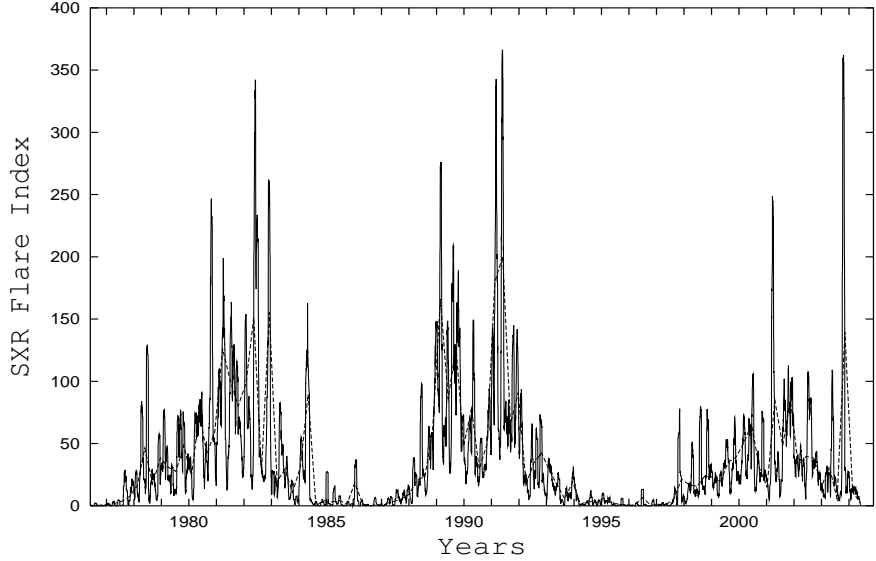,width=7.5cm, height=12cm,angle=270}
\caption{ 27-day moving average of daily $FI_{SXR}$ for the period 1976$-$2004.
The dashed line shows the smooth spline curve.}
\end{figure} 
     
\begin{figure}
\epsfig{file=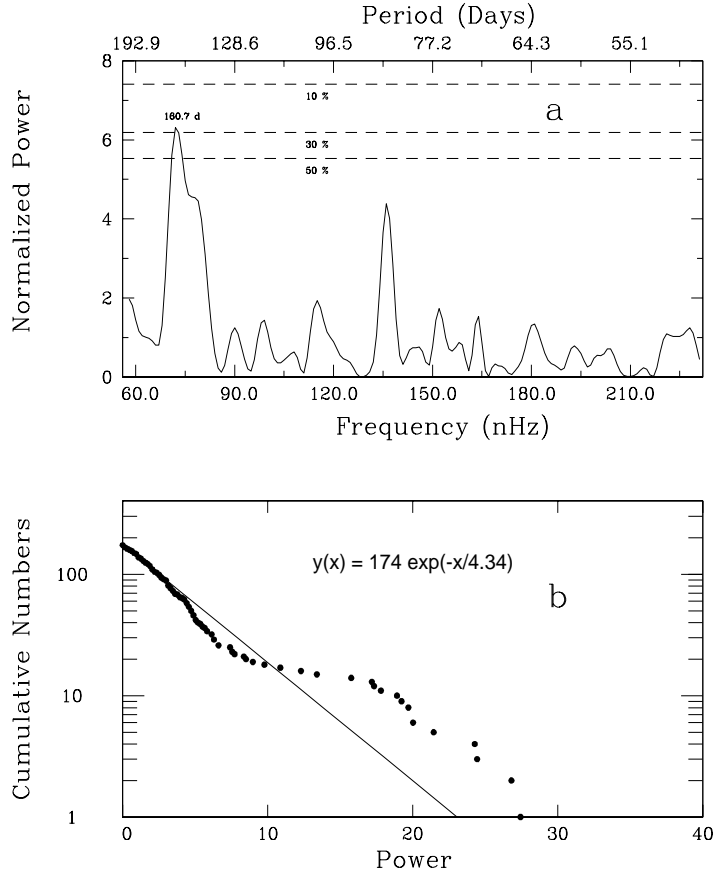,width=11.5cm, height=12cm}
\caption{(a) Normalized power spectrum of the daily $FI_{SXR}$ time series
of solar cycle 21 for the frequency interval of 231.5$-$57.8 nHz 
(50$-$200 days). Horizontal dashed lines indicate FAP levels of 10 $\%$, 30 $\%$, and 
50 $\%$ (from top to bottom) (b) Scargle
power distribution corresponding to (a). The vertical
axis is the number of frequencies for which power exceeds x. The 
straight line is the fit to the points for lower values of power.}
\end{figure}      

\begin{figure}
\epsfig{file=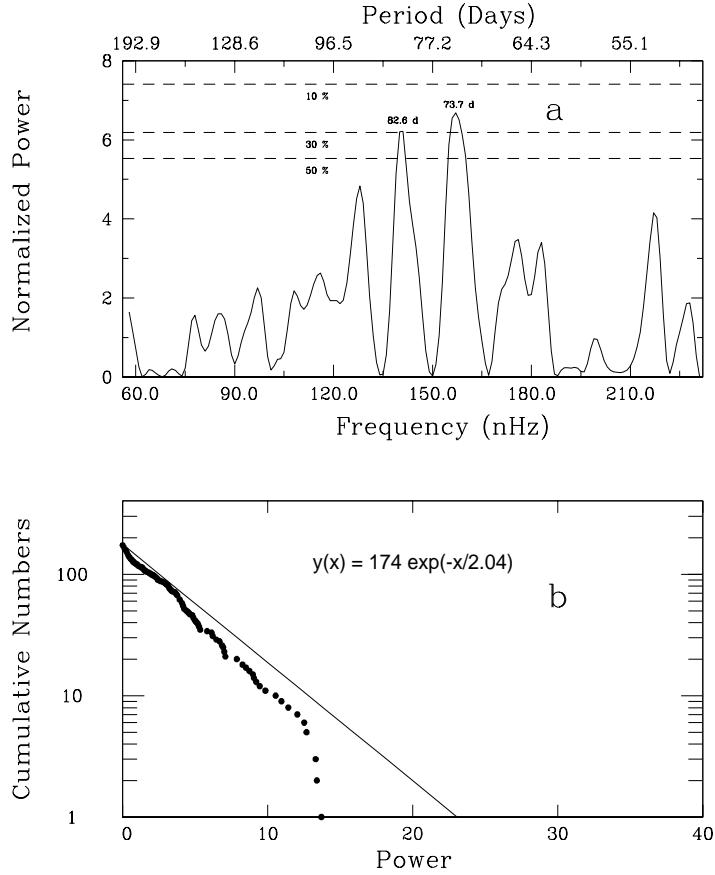, width=11.5cm, height=12cm}
\caption{Same as Figure 1, but for solar cycle 22.}
\end{figure}

\section{Periodogram Analysis }
We have Used the Lomb-Scargle periodogram method (Lomb, 1976; Scargle, 
1982) modified by Horne and Baliunas (1986) to perform the power spectra
of $FI_{SXR}$ time series. The power spectra for the three cycles
21, 22 and, 23 has been computed for the frequency range of 231.5$-$57.8 nHz 
which corresponds to the period interval of 50$-$200 days.
Figure 2(a) shows the normalized 
power spectra of daily $FI_{SXR}$ values for solar cycle 21. 
The $FI_{SXR}$ is not 
independent. Therefore the probability P of the power density at a given 
frequency being greater than K by chance is given by
\begin{equation}
P(z > K) = exp (-K/k),
\end{equation}
where the normalization factor $k$, which is due to event correlation,
can be determined empirically (Bai and Cliver, 1990).

\begin{figure}
\epsfig{file=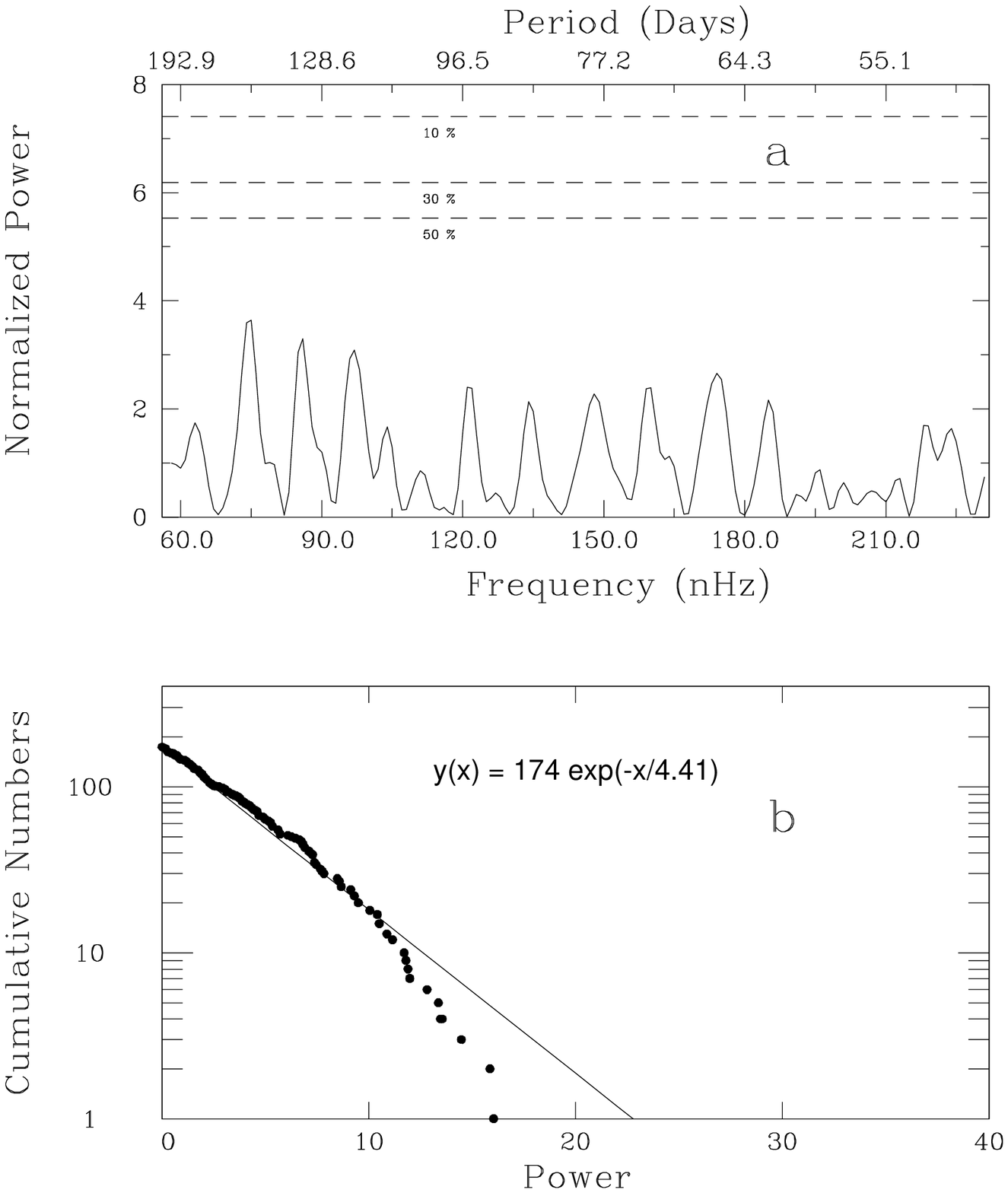, width=11.5cm, height=12cm}
\caption{Same as Figure 1, but for solar cycle 23.}
\end{figure}

\begin{figure}
\epsfig{file=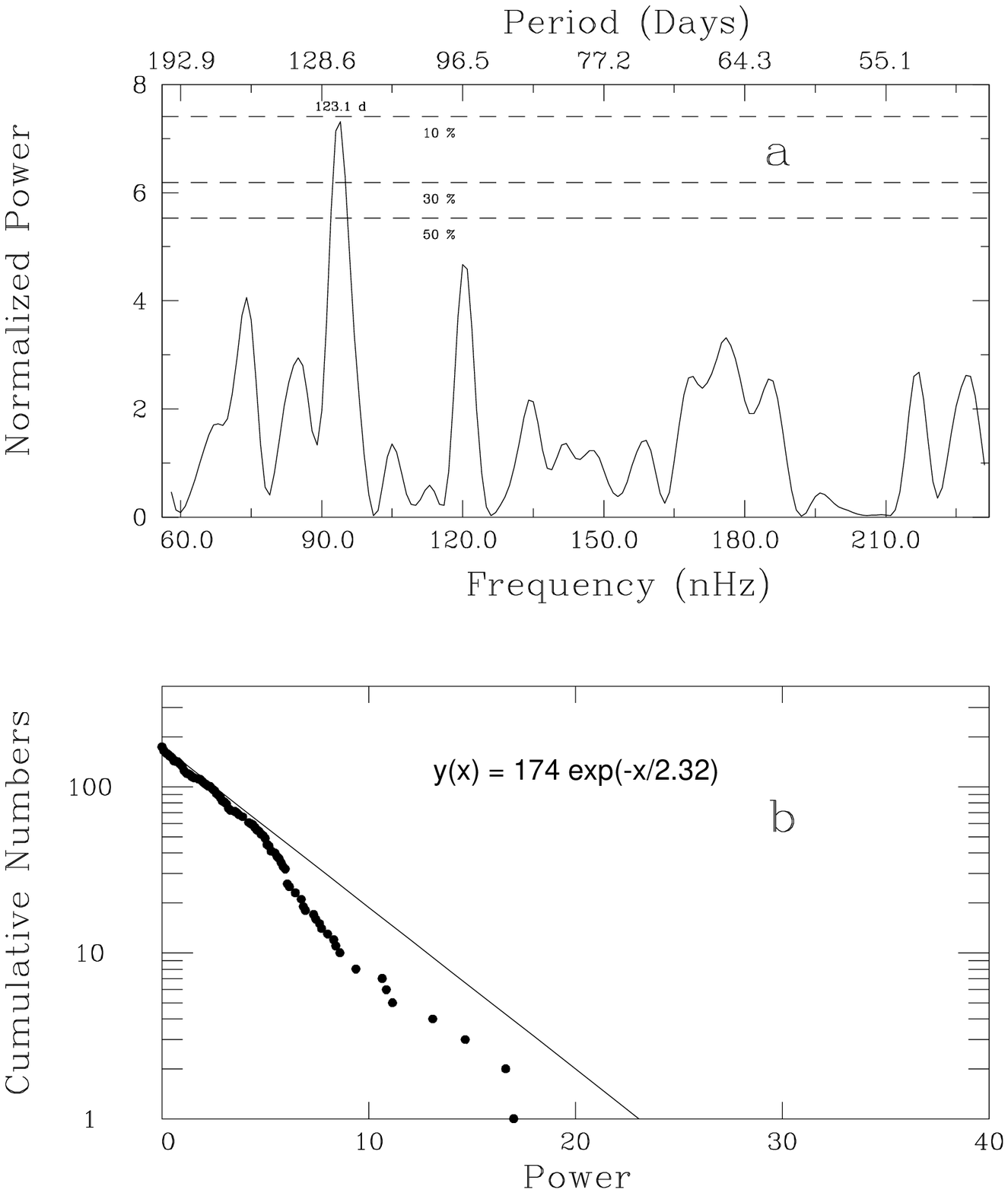, width=11.5cm, height=12cm}
\caption{Same as Figure 1, but for the interval from May 1996 to October 2002 
of 
the solar cycle 23.}
\end{figure}

Figure 2(b) shows the distribution of the power values
corresponding to the normalized power spectra shown in Figure
2(a). The vertical axis shows the cumulative 
number of frequencies
for which the power exceeds a certain value. For example, in the 
power spectrum for solar cycle 21 shown in Figure 2(a) we have 174
frequencies. For all these frequencies the power exceeds zero; thus, 
we have a point at (X=0, Y=174). At only one frequency 
(72.0 nHz, which is equivalent to 160.75 days) the power was 27.42,
its maximum value. For  lower values of power, the distribution
can be well fitted by the equation y = 174 exp(-x/4.34), as expected
from equation (1). Thus, we normalize the power spectrum by 
dividing the powers by 4.34 to obtain Figure 2(a). For other cases we
use the same procedure for normalization.

Once the power spectrum has been normalized properly, we can use the
`False alarm probability' (FAP) formula to estimate the statistical
significance of a peak in the power spectrum. It is given by the 
expression
\begin{equation}
F = 1 - [1 - exp(-Z_{m})]^{N},
\end{equation}
where $Z_{m}$ is the height of the peak in the normalized power
spectrum and $N$ is the number of independent frequencies.
The interpretation of $F$ is as follows: if we have
a discrete power spectrum giving the power at each of $N$ independent frequencies
for a set of random data, $F$ indicates the probability that the power at one or more
of these frequencies will exceed $Z_{m}$ by chance. Fourier components calculated
at frequencies at intervals of independent fourier spacing, $\Delta$$f_{ifs}$=$\tau$$^{-1}$, where
$\tau$ is the time span of the data, are totally independent (Scargle, 1982). For example, in the
power spectrum shown in Figure 2(a), we have $\tau$=3717 days and $\Delta$$f_{ifs}$=3.1 nHz. 
Thus there are 56 independent frequencies in 231.5$-$57.8 nHz interval. We have oversampled
to obtain this power spectrum in which the height of the peak at 160.75 days is 6.3. The 
oversampling tends to estimate more accurately the peak value.    
Therefore, if we substitute $Z_{m}$=6.3
and N=174 (since we searched 174 frequencies with 1.0 nHz
intervals) in equation (2) we get $F$=0.27, i.e., the probability
to obtain such a high peak at 72.0 nHz (160.75 days) by chance is about 27$\%$.
The same analysis has been applied to the other power spectra also  shown in 
Figure (3), (4), and (5). In the power spectra for solar cycle 22 (Figure 3)
we have found two important peaks at 157.0 nHz (73.72 days) with FAP=19 $\%$
and 140.4 nHz (82.67 days) with FAP=32.9 $\%$. No significant peak appears
in the power spectra for solar cycle 23 (Figure 4) when the data  
from the beginning of cycle 23 to 31 May 2004 was analyzed. We have also
performed the power spectra for three different time intervals (May 1996 to
October 2002, September 1999 to June 2001 and October 1999 to April 2001)
studied by Bai (2003). We could not find any statistically significant
peak in the second and third duration. However in the first time interval
(i.e., May 1996 to October 2002) a significant peak appears at 94.0 nHz 
(123.13 days) with FAP=11$\%$ (Figure 5).

To be sure that peaks in the different power spectra (Figures 2, 3, and 5)
are not due to aliasing, we have removed sine curves of their periods
from the original time series. We see that these peaks get removed
in the power spectra of time series obtained after subtraction.

\section{Summary and Discussion}
We have studied periodicities in $FI_{SXR}$ during solar cycles 21, 22 and 23 
separately. It will be useful to compare and discuss our results in the light
of periodicities detected in other solar activity indicators. 

We have found that a period of 161 days appeared in $FI_{SXR}$ during solar
cycle 21, which is absent during cycle 22 and 23.
The near 155 days periodicity, first detected by Rieger et al. (1984) in
gamma ray and soft X-ray flares during cycle 21, has been found
in several other studies. 
During
the same solar cycle, it was detected in flare producing energetic 
interplanetary electrons (Dr\"{o}ge et al., 1990), proton flares (Bai and 
Cliver,
1990), and microwave flares (Bogart and Bai, 1985; Kile and Cliver, 1991). This 
periodicity was also found in H$\alpha$ flares (Ichimoto et al., 1985), 
microwave flares (Bogart and Bai, 1985), 
and H$\alpha$ flare
index (\"{O}zg\"{u}c and Ata\c{c}, 1989) when the data of solar cycles 20 and 21 was 
considered together. 
Lean and Brueckner (1989) detected a periodicity at 155 days in the sunspot
blocking function and the 10.7 cm radio flux during solar cycles 19, 20
and 21.
Carbonell and Ballester (1990, 1992) have shown that a periodicity 
around 150$-$160 days seems to be significant during all solar cycles 
from 16 to 21 in sunspot areas.  
Recently Ballester, Oliver, and Carbonell (2002) in the 
analysis of historical record of photosphere magnetic flux detected a 
significant periodicity near 160 days during solar cycle 21.

Several mechanism
have been put forward to explain the cause of near 155 days periodicity.  
Bai and Sturrock (1991) and Sturrock and Bai (1992) proposed that Sun 
contains a ``clock" with a period of 25.8 days (later modified as 25.5 days)
and suggested that periodicity
of 154 days is a subharmonic of that fundamental period. Bai and Sturrock
(1993) analyzed the longitude distribution of major flares of cycles 19$-$22
and interpreted this hypothetical clock as being an obliquely rotating 
structure (or a wave pattern) rotating with a period of 25.5 days 
about an axis titled by 40$^{o}$ with respect to the solar rotation axis.    
On the other hand, another explanation suggests that this periodicity  
is associated
with the periodic emergence of the magnetic flux linked to the regions
of strong magnetic field (Carbonell and Ballester, 1990, 1992; 
Oliver, Ballester, and Baudin, 1998; Ballester, Oliver, and Baudin, 1999;
Ballester, Oliver, and Carbonell, 2002). 

Other periodicities detected by us in $FI_{SXR}$ during cycles 22 and 23
are also well consistent with the earlier work. We detected 74 days 
and 83 days periodicities in 
$FI_{SXR}$ for cycle 22. Bai (1992) analyzed the major flare 
occurrence of solar cycle 22 and found a periodicity of 77 days 
which was interpreted as the third subharmonic of the 25.5 days period. 
\"{O}zg\"{u}c and Ata\c{c} (1994) also 
found a period of 73 days in $H\alpha$ flare index during cycle 22. Oliver
and Ballester (1995), in the analysis of sunspot areas, found a periodicity
around 86 days (close to 83 days period in $FI_{SXR}$) which is statistically
significant during part of solar cycle 22. For cycle 23 we have carried out
analysis in four different time intervals, but a significant peak appears 
only in the duration of May 1996 to October 2002 with a period of 123 days.
Lou et al. (2003) found a periodicity of 122 days in X-ray flares 
of class $\geq$ M5.0
during the maximum phase of solar cycle 23.
Bai (2003) detected a periodicity of 129 days in major flares during part of
cycle 23 (September 1999 to June 2001). Here it is interesting to note that 
74 and 123 days periodicities, found in $FI_{SXR}$,
are close to integral multiple
of 25.5 days.

\begin{acknowledgements}
We are thankful to Prof. Ram Sagar for valuable suggestions.
One of the authors (BJ) wishes to thank Dr. Brijesh Kumar for useful 
discussions. We thank an anonymous referee for helpful comments and suggestions.
BJ also wishes to thank Dr. P. Pant for several useful discussions.
\end{acknowledgements}

\end{article}

\addcontentsline{toc}{section}{References}
\begin{thebibliography}{}  


\bibitem[\protect\citeauthoryear{AI}{1992}]{Bai1992}
Bai, T.: 1992, {\it Astrophys. J.}
{\bf 388}, L69.

\bibitem[\protect\citeauthoryear{Bai}{2003}]{Bai2003}
Bai, T.: 2003, {\it Astrophys. J.}
{\bf 591}, 406.

\bibitem[\protect\citeauthoryear{Bai}{1990}]{Bai90}
Bai, T. and Cliver, E. W.: 1990, {\it Astrophys. J.}
{\bf 363}, 299.

\bibitem[\protect\citeauthoryear{Bai}{2003}]{Bai2003}
Bai, T. and Sturrock, P. A.: 1991, {\it Nature}
{\bf 350}, 141.

\bibitem[\protect\citeauthoryear{Bai}{1993}]{Bai1993}
Bai, T. and Sturrock, P. A.: 1993, {\it Astrophys. J.}
{\bf 409}, 476.

\bibitem[\protect\citeauthoryear{}{}]{}
Ballester, J. L., Oliver, R., and Baudin, F.: 1999, {\it Astrophys. J.} 
{\bf 522}, L153.       

\bibitem[\protect\citeauthoryear{}{}]{}
Ballester, J. L., Oliver, R., and Carbonell, M.: 2002, {\it Astrophys. J.} 
{\bf 566}, 505.       

\bibitem[\protect\citeauthoryear{Bogart}{1985}]{Bogart85}
Bogart, R. S. and Bai, T.: 1985 {\it Astrophys. J.}
{\bf 299}, L51.

\bibitem[\protect\citeauthoryear{}{}]{}
Carbonell, M. and Ballester, J. L.: 1990, {\it Astron. Astrophys.} 
{\bf 238}, 377.       

\bibitem[\protect\citeauthoryear{}{}]{}
Carbonell, M. and Ballester, J. L.: 1992, {\it Astron. Astrophys.} 
{\bf 255}, 350.       

\bibitem[\protect\citeauthoryear{Droge}{1990}]{Droge1990}
Dr\"{o}ge, W., Gibbs, K., Grunsfeld, J. M., Meyer, P., Newport, B. J., 
Evenson, P., and Moses, D.: 1990, {\it Astrophys. J. Supll. Ser.}
{\bf 73}, 279.

\bibitem[\protect\citeauthoryear{Horne}{1986}]{Horne86}
Horne, J. H. and Baliunas, S. L.: 1986, {\it Astrophys. J.}
{\bf 302}, 757.

\bibitem[\protect\citeauthoryear{Ichimoto}{1985}]{Ichimoto85}
Ichimoto, K., Kubota, J., Suzuki, M., Tohmura, I., and Kurokawa, H.: 1985, {\it Nature}
{\bf 316}, 422.

\bibitem[\protect\citeauthoryear{Joshi}{2004}]{Joshi04}
Joshi, B. and Joshi, A.: 2004, {\it Solar Phys.}
{\bf 219}, 343.

\bibitem[\protect\citeauthoryear{Kile}{1985}]{Kile85}
Kile, J. N. and Cliver, E. W.: 1991, {\it Astrophys. J.}
{\bf 370}, 442.

\bibitem[\protect\citeauthoryear{Lean}{1990}]{Lean90}
Lean, J. L.: 1990 {\it Astrophys. J.}
{\bf 363}, 718.

\bibitem[\protect\citeauthoryear{Lean}{1989}]{Lean89}
Lean, J. L. and Brueckner, G. E.: 1989 {\it Astrophys. J.}
{\bf 337}, 568.

\bibitem[\protect\citeauthoryear{Lomb}{1976}]{Lomb76}
Lomb, N.: 1976, {\it Astrophys. Space Sci.}
{\bf 39}, 477.

\bibitem[\protect\citeauthoryear{Lou}{2003}]{Lomb76}
Lou, Y., Wang, Y., Fan, Z., Wang, S., and Wang, J.: 2003, {\it Monthly Notices 
Royal Astron. Soc.}
{\bf 345}, 809.

\bibitem[\protect\citeauthoryear{Oliver et al.}{1995}]{Oliver95}
Oliver, R., Ballester, J.L.: 1995, {\it Solar Phys.}
{\bf 156}, 145.

\bibitem[\protect\citeauthoryear{Oliver et al.}{1998}]{Oliver98}
Oliver, R., Ballester, J. L., and Baudin, F.: 1998, {\it Nature}
{\bf 394 }, 552.

\bibitem[\protect\citeauthoryear{Ozguc et al.}{1989}]{Ozguc89}
\"{O}zg\"{u}\c{c}, A. and Ata\c{c}, T.: 1989, {\it Solar Phys.}
{\bf 123}, 357.

\bibitem[\protect\citeauthoryear{Ozguc et al.}{1994}]{Ozguc94}
\"{O}zg\"{u}\c{c}, A. and Ata\c{c}, T.: 1994, {\it Solar Phys.}
{\bf 150}, 339.

\bibitem[\protect\citeauthoryear{Rieger}{1984}]{Rieger84}
Rieger, E., Share, G. H., Forrest, D. J., Kanbach, G., Reppin, C., and Chupp, 
E. L.: 1984, {\it Nature}
{\bf 312}, 623.

\bibitem[\protect\citeauthoryear{Scargle}{1982}]{Scargle82}
Scargle. J. D.: 1982, {\it Astrophys. J.}
{\bf 263}, 835.


\bibitem[\protect\citeauthoryear{Sturrock}{1992}]{Sturrock92}
Sturrock, P. A. and Bai, T.: 1992, {\it Astrophys. J.}
{\bf 397}, 337.

\end{thebibliography}
\end{document}